\documentclass[12pt]{article}

\usepackage[body={17cm, 22.5cm},right=2.2cm]{geometry}

\usepackage{booktabs}
\usepackage{caption}

\usepackage{color}
\usepackage{graphicx}
\usepackage{epsf}
\usepackage{graphicx,epsfig}
\usepackage{multirow}

\usepackage{amsmath, amsthm, amssymb}

\usepackage{epsfig}
\usepackage{cite}
\usepackage{color,colordvi}

\newcommand{\be}{\begin{eqnarray}}
\newcommand{\ee}{\end{eqnarray}}
\newcommand{\bi}{\begin{itemize}}
\newcommand{\ei}{\end{itemize}}

\def\<{\left\langle}
\def\>{\right\rangle}

\def\nn{\nonumber}
\def\p{\partial}
\def\ls{\left[}
\def\rs{\right]}
\def\lc{\left\{}
\def\rc{\right\}}

\newcounter{hran}

\makeatletter
\renewcommand\section{\@startsection {section}{1}{\z@}%
                               {-3.5ex \@plus -1ex \@minus -.2ex}%
                               {2.3ex \@plus.2ex}%
                               {\normalfont\large\bfseries}}
\makeatother

\begin{document}

\vspace{5mm}

\begin{center}
\vglue .10in
{\Large\bf {  Starobinsky inflation from new-minimal supergravity  }
}

{\ }
\\[0.3cm]
{\large  Fotis Farakos}
\\[0.2cm]

\vspace{.3cm}
{\normalsize {\it  Institute for Theoretical Physics, Masaryk University, \\611 37 Brno, Czech Republic}}

\vspace{.3cm}
{\normalsize  E-mail: fotisf@mail.muni.cz }

\end{center}

\noindent
{\it Proceedings for DISCRETE 2014 
\\
2-6 December 2014 
\\
King's College London, Strand Campus  }

\vspace{.4cm}

{\small  \noindent \textbf{Abstract} \\[0.3cm] 
In the  new-minimal supergravity formulation we present the embedding of the $R+R^2$ Starobinsky model of inflation. 
Starting from the superspace action we perform the projection to component fields 
and identify the Starobinsky model in the bosonic sector. 
Since there exist no other scalar fields, 
this is by construction a single field model. 
This higher curvature supergravity also gives rise to a propagating  massive vector. 
Finally we comment on the issues of higher order corrections and initial conditions.

\vskip 0.5cm

\def\thefootnote{\arabic{footnote}}
\setcounter{footnote}{0}

%
%
%
%

\def\thefootnote{\arabic{footnote}}
\setcounter{footnote}{0}

\def\ls{\left[}
\def\rs{\right]}
\def\lc{\left\{}
\def\rc{\right\}}

\def\p{\partial}

\def\S{\Sigma}

\def\nn{\nonumber}

\section{Introduction and discussion}

The Planck collaboration's constraints on inflation 
\cite{Ade:2013uln,Ade:2015lrj} have  restricted the inflationary models 
to those which are characterized by a plateau potential for the inflaton field. 
More specifically, if the perturbations during inflation \cite{Lyth:1998xn} are originated by the same field driving inflation, 
these restrictions can be quantified by the following constraints on the spectral index: $n_s = 0.9655 \pm 0.0062$  
and the tensor-to-scalar ratio: $r < 0.12$.  
A model which lies in the heart of the data is the pure gravitational Starobinsky model  \cite{Starobinsky:1980te} 
\be
e^{-1} {\cal L}= \frac{M_P^2}{2}  R + \frac{M_P^2}{12 m^2} R^2 , 
\ee
which in the Einstein frame describes a real scalar field minimally coupled to gravitation, 
with a scalar potential given by 
\be
V_{R^2} = \frac34 m^2 M_P^2 \left( 1 - e^{- \sqrt{\frac23} \phi / M_P} \right)^2 . 
\ee 
This model gives $n_s  -1 \simeq -2/N  $ and $r \simeq 12/N^2$. 
The Planck data constrain $m \simeq 1.3 \times 10^{-5} M_P$. 
Discussions on the generic properties of models with plateau 
potentials can be found in \cite{Kehagias:2013mya,Biagetti:2015tja}.

Supergravity \cite{Ferrara:1988qxa,Wess:1992cp}, as the low energy limit of string theory, 
is essentially the appropriate framework to study high energy gravitational phenomena like inflation. 
The minimal 4D N=1 supergravity multiplet contains as physical fields the graviton, 
with 6 bosonic off-shell degrees of freedom, 
and the gravitino, 
with 12 off-shell fermionic degrees of freedom. 
The remaining  6 off-shell bosonic degrees of freedom are auxiliary and can be distributed as follows 
\begin{itemize} 

\item Old-minimal supergravity  auxiliary fields sector \cite{Ferrara:1978em,Stelle:1978ye}: 
a complex scalar $M$ (2 DOF) and a real vector $b_m$ (4 DOF).

\item New-minimal supergravity  auxiliary fields sector \cite{Sohnius:1981tp}: 
a gauge vector for the R-symmetry $A_m$ (3 DOF) and a gauge two-form $B_{mn}$ (3 DOF). 

\end{itemize}
The existence of different minimal supergravities can be understood as arising from different solutions to the superspace 
Bianchi identities, or as different choice of appropriate Wess-Zumino gauge for the gravitational multiplet, 
or also as  originating from the different compensating multiplets that break the 
underlying superconformal theory to super-Poincar\'e. 
The underlying dualities among the compensating multiplets 
survive the gauge fixing and lead to equivalent couplings to matter \cite{Ferrara:1983dh}, 
but break down as soon as higher curvature terms  are introduced. 
Here we  present the embedding  of the  
Starobinsky model of inflation  in new-minimal supergravity \cite{Farakos:2013cqa}. 
This higher curvature supergravity  is on-shell equivalent to standard supergravity coupled to 
a massive vector multiplet \cite{Cecotti:1987qe,Farakos:2013cqa,Ferrara:2013rsa,Ferrara:2014cca}.

The Starobinsky model of inflation is nevertheless accompanied by a series of open issues. 
The first concerns the existence of possible higher order curvature corrections. 
For example the $R^4$ terms \cite{Farakos:2013cqa}. 
As we will see, 
these terms spoil the plateau of the scalar potential and if they are relatively large   Starobinsky inflation does not take place. 
Therefore, one requires a hierarchy to hold during inflation 
\be
\label{hir}
\frac{M_P^2}{m^2} R_{inf}^2 \gg M_P^2 R_{inf} \ \ , \ \ \frac{M_P^2}{m^2} R_{inf}^2 \gg \xi R_{inf}^4 \, ,   
\ee
where $\xi$ is an appropriate parameter for the $R^4$ terms. 
This hierarchy has no apparent justification and 
a concrete answer to why the $\xi R^4$ terms are expected to be small (therefore pose no threat) is not known.  
Proposals of why the hierarchy (\ref{hir}) is expected to hold 
in a supergravity framework  can be found in \cite{LopesCardoso:1992wv,Ferrara:2013kca}.

A second open issue, which is again related to the scale $m \sim 10^{-5} M_P$ is the initial conditions problem  \cite{Goldwirth:1991rj,Ijjas:2013vea}. 
If our universe exited the quantum gravity regime with an 
energy density $\sim M_P^4$ \cite{Linde:2005ht,Linde:2007fr}, 
then due to the characteristic upper bound of the potential energy of the Starobinsky model $\sim 10^{-10} M_P^4$, 
the total energy density has to be dominated by the kinematic contribution. 
This leads to a need for an initial homogeneous patch of radius of the order 
\be
r_{init} \sim 10^3 \, l_P,
\ee 
which means rather special initial conditions. 
A proposal of how this is ameliorated in a pure $R+R^2$ setup (supergravity or not), 
has been given in \cite{Dalianis:2015fpa} which we also review here. 
For a review on inflationary cosmology after the release of the Planck collaboration's results, 
and for an approach on the initial conditions problem see also \cite{Linde:2014nna}.

Note that the embedding of the Starobinsky model has been also studied 
in the old-minimal supergravity  \cite{Ferrara:1978rk,Cecotti:1987sa,Kallosh:2013lkr,Farakos:2013cqa,Ferrara:2013wka,Dalianis:2014aya,Terada:2014uia}, 
in no-scale supergravity \cite{Ellis:2013xoa,Ellis:2013nxa,Ellis:2014gxa,Lahanas:2015jwa}, 
in the linearized non-minimal (20/20) supergravity \cite{Farakos:2015hfa}, 
and also in a generic supergravity setup via gravitino condensates \cite{Alexandre:2013nqa,Alexandre:2014lla}.

\section{$R+R^2$ new-minimal supergravity}

The  new-minimal supergravity  \cite{Sohnius:1981tp} is the supersymmetric theory of the gravitational multiplet
\begin{equation}
e^a_m\, ,~~\psi_m^\alpha \,  , ~~A_m\, , ~~ B_{mn}\ .
\end{equation}
The first two fields are the vierbein and its superpartner the gravitino, 
a spin-$\frac{3}{2}$ Rarita-Schwinger field. 
The last two fields are auxiliaries. 
The real auxiliary vector $A_m$ gauges the $U(1)_\text{R}$ chiral symmetry. 
The auxiliary $B_{mn}$ is a real two-form appearing only through its dual field strength $H_m$, 
which satisfies $\hat D^a H_a =0$ for the supercovariant derivative $\hat D^a$.

We will use superspace techniques to guarantee that our component form 
Lagrangians are supersymmetric.  
The interested reader may consult for example \cite{Ferrara:1988qxa} where a treatment of 
the new-minimal superspace is given. 
The new-minimal supergravity free Lagrangian is given by
\begin{equation}
\label{sugra} 
{\cal L}_0= - 2 M^2_P  \int d^4 \theta E V_{\text{R}} . 
\end{equation}
Here $V_{\text{R}}$ is the gauge multiplet of the R-symmetry, 
which (in the appropriate WZ gauge) contains the auxiliary fields in its vector component
\be
- \frac{1}{2} [\nabla_\alpha , \bar \nabla_{\dot \alpha} ]  V_{\text{R}} | = A^-_{\alpha \dot \alpha} 
= A_{\alpha \dot \alpha} - 3 H_{\alpha \dot \alpha} ,
\ee
and the Ricci scalar in its highest component 
\be
\frac{1}{8} \nabla^\alpha \bar \nabla^2 \nabla_\alpha  V_{\text{R}} |= - \frac{1}{2} \left( R + 6 H^a H_a \right) . 
\ee
The symbol $E$  stands for the super-determinant of new-minimal supergravity.  
The bosonic sector of Lagrangian (\ref{sugra}) is
\begin{equation}
\label{freesugra}
{\cal L}_0= 
  M^2_P \, e \, \left( \frac{1}{2}
R + 2A_a H^a-3H_a H^a\right) .
\end{equation}

It is well known from linearized supergarvity \cite{Cecotti:1987qe} that the $R^2$ term is accommodated inside
\be
\label{R2}
{\cal L}_{R^2} =  \frac{\alpha}{4} \int d^2 \theta  \, {\cal E} W^2(V_{\text{R}}) + c.c.
\ee
with the standard definition of the field strength 
\be
W_{\alpha}(V_{\text{R}}) = -\frac{1}{4} \bar \nabla^2 \nabla_\alpha V_{\text{R}}, 
\ee
and ${\cal E}$ the chiral density. 
In component form, the bosonic sector of  (\ref{R2}) reads  
\be
e^{-1} {\cal L}_{R^2} = \frac{\alpha}{8} \left(R+6H^2 \right)^2  - \frac{\alpha}{4} F^2(A^-) .
\ee
The Starobinsky model of inflation in new-minimal supergravity (now we set $M_P=1$) is \cite{Farakos:2013cqa} 
\be
\nn
{\cal L} = -2 \int d^4 \theta \, E V_{\text{R}} + \frac{\alpha}{4} \int d^2 \theta \, {\cal E} W^2(V_{\text{R}}) + c.c.
\ee
with bosonic sector
\be
\label{star}
e^{-1} {\cal L}=   \frac{1}{2} R+ 2A_a H^a-3H_a H^a 
+ \frac{\alpha}{8} \left(R+6H^2\right)^2  - \frac{\alpha}{4} F^2(A^-) .
\ee
Indeed, theory (\ref{star}) describes $R + R^2$, but the curvature terms are mixed with the auxiliary field $H^a$. 

To find the theory in the Einstein frame we proceed to integrating out the auxiliary fields. 
The Lagrangian (\ref{star}) is classically equivalent to 
\be
e^{-1} {\cal L}=  \frac{1}{2} R+ 2A_a H^a - 3H_a H^a - 2 H^m \p_m X 
+ \frac{\alpha}{8}  \Psi   \left(R+6H^2\right) -  \frac{\alpha}{32}  \Psi^2    - \frac{\alpha}{4} F^2(A^-) ,
\ee
where now $H_m$ is an unconstrained real vector. 
Indeed, by integrating out  the real scalar  $X$, 
it imposes the appropriate constraint on $H_m$, 
and by integrating out $Y$ we get (\ref{star}). 
Now we redefine the auxiliary $A_m$ as
\be
{\cal V}_m = A_m -3 H_m - \p_m X,
\ee 
and we find
\be
e^{-1} {\cal L}=  \frac{1}{2}\left(1 +  \frac{\alpha}{4}  \Psi  \right)  R   - \frac{\alpha}{4} F^2({\cal V})  
 + 2{\cal V}_a H^a + 3 \left(1 +  \frac{\alpha}{4}  \Psi  \right) H^2  -  \frac{\alpha}{32}  \Psi^2  .
\ee
The auxiliary $H_m$ has become quadratic and after it is integrated out we have  
\be
e^{-1} {\cal L}=  \frac{1}{2}\left(1 +  \frac{\alpha}{4}  \Psi  \right)  R   - \frac{\alpha}{4} F^2({\cal V})  
 -  \frac{\alpha}{32}  \Psi^2  - \frac{1}{3} \frac{{\cal V}^2}{\left(1 +  \frac{\alpha}{4}  \Psi  \right)} .
\ee
Note that the original $A_m$ not only has become propagating, but it has also become massive. 
After rescaling the theory to go to the Einstein frame by 
a conformal transformation
\be
e_m^{\ a} \rightarrow \frac{1}{ \sqrt{1 +  \frac{\alpha}{4}  \Psi }} \, e_m^{\ a} ,
\ee
we have (for $\Psi \rightarrow \Psi / \alpha$ and ${\cal V} \rightarrow {\cal V} / \sqrt \alpha$)
\be
e^{-1} {\cal L}=  \frac{1}{2}  R   - \frac{1}{4} F^2({\cal V})  -\frac{3}{64 (1 +  \frac{1}{4}  \Psi )^2} \p \Psi \p \Psi
 -  \frac{1}{32 \alpha}  \Psi^2 \frac{1}{ (1 +  \frac{1}{4}  \Psi )^2}
 - \frac{1}{3 \alpha} \frac{{\cal V}^2}{\left(1 +  \frac{1}{4}  \Psi  \right)^2} .
\ee
Finally, for 
\be
\phi = \sqrt{\frac32} \, \text{ln} \left(1 +  \frac{1}{4}  \Psi \right),
\ee
we have (for $\frac1\alpha = 9 g^2$) 
\be
\label{tot}
e^{-1} {\cal L}=  \frac{1}{2}  R   - \frac{1}{4} F^2({\cal V})  - \frac12  \p \phi \, \p \phi
 -  \frac{9 g^2}{2} \left( 1 - e^{- \sqrt{\frac23} \phi}  \right)^2
 - 3 g^2 e^{- 2 \sqrt{\frac23} \phi} \, {\cal V}^2.
\ee
This is the  dual form of the Starobinsky model. 
Here we have reproduced it in a new-minimal supergravity framework \cite{Farakos:2013cqa}.  
Notice that there is only one real propagating scalar field, and a massive vector.  
From (\ref{tot}) one can calculate the slow-roll parameters and verify that the model is in perfect agreement 
with the Planck data \cite{Ade:2013uln}.

\section{Open issues in the Starobinsky model}

In this part we review some known open issues of the Starobinsky model.

\subsection{Higher order corrections}

On  top of the $R+R^2$ theory one could ask what is the impact of the higher order corrections. 
We will consider here the $R^4$ terms. 
The superspace Lagrangian for $R^4$ has the form \cite{Farakos:2013cqa} 
\be
{\cal L}_{R^4} = 16 \xi \int d^4 \theta E \, W^2(V_{\text{R}}) \bar W^2(V_{\text{R}}) . 
\ee
The full Lagrangian including the $R^4$ terms reads 
\be
{\cal L} = -2 \int d^4 \theta \, E V_{\text{R}} 
+ \lc \frac{\alpha}{4} \int d^2 \theta \, {\cal E} W^2(V_{\text{R}}) + c.c. \rc
+ 16 \xi \int d^4 \theta E \, W^2(V_{\text{R}}) \bar W^2(V_{\text{R}}) . 
\ee

During inflation only the curvature terms contribute, 
therefore we can work with the Lagrangian
\be
\label{r4}
e^{-1} {\cal L} = \frac12 R + \frac{\alpha}{8} R^2 +  \xi R^4 .
\ee
The bosonic terms that we have ignored in writing (\ref{r4}) would only contribute 
to the vector sector in the dual description (see \cite{Farakos:2013cqa,Ferrara:2013rsa,Ferrara:2013kca}).  
We can then rewrite the theory with the use of a Lagrange multiplier $Z$ as 
\be
e^{-1} {\cal L} = (\frac12 + Z) R + \frac{\alpha}{8} Y^2 +  \xi Y^4 - Z Y. 
\ee
Indeed, by integrating out $Z$ we find $Y=R$ and we get (\ref{r4}). 
Now we proceed in the other direction and we integrate out $Y$. 
The equation of motion for $Y$ gives 
\be
Y^3 + \frac{\alpha}{16 \xi} Y - \frac{Z}{4 \xi} = 0, 
\ee
which can be solved as
\be
Y(Z) = \frac13 \left( \frac{27}{8 \xi} Z + \frac12 \sqrt{\left( \frac{27}{4 \xi} Z \right)^2  
+ 4 \left( \frac{3 \alpha}{16 \xi} \right)^3 } \right)^{\frac13} 
- \frac{\alpha}{16 \xi}   \left( \frac{27}{8 \xi} Z + \frac12 \sqrt{\left( \frac{27}{4 \xi} Z \right)^2  
+ 4 \left( \frac{3 \alpha}{16 \xi} \right)^3 } \right)^{-\frac13} .  
\ee
After integrating out $Y$,  rescaling the metric and redefining $Z$ we find 
\be
e^{-1} {\cal L} = \frac12 R - \frac12 \p \phi \p \phi - V(\phi),  
\ee
with scalar potential
\be
\label{R4}
V(\phi) =  \frac{3 \xi Y^4(Z) + \frac{\alpha}{8} Y^2(Z) }{4 (Z+\frac12)^2} \Big{|}_{Z = \frac12 e^{\sqrt{\frac23} \phi} -\frac12} . 
\ee
The plot of the scalar potential (\ref{R4}) can be seen in  Figure 1. 
It is easy to see that for small $\xi$ values inflation is not ruined, 
but larger $\xi$ values may  pose a threat \cite{Farakos:2013cqa,Ferrara:2013kca}.

\begin{figure}[htp] \centering{
\includegraphics[scale=1.2]{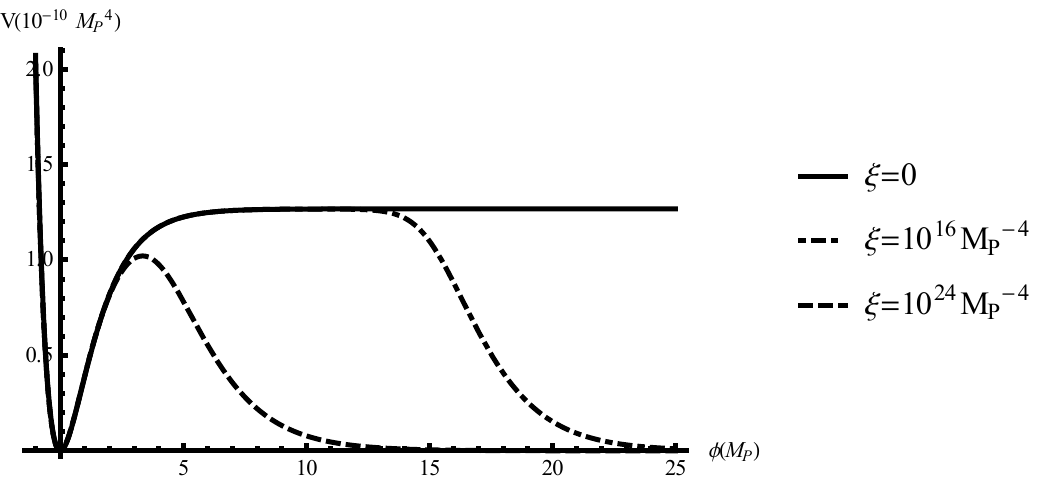}}
\caption{The potential for the Starobinsky model in the dual description, including $R^4$ terms parameterized by $\xi$. 
Here we have set $\alpha \sim 0.4 \times 10^{10} M_P^2$ as constrained by the Planck data. 
One can see the characteristic plateau of the Starobinsky model (for $\xi=0$) at $V_{inf} \sim 1.3 \times 10^{-10} M_P^4.$ }
\end{figure}

\subsection{Initial conditions problem}

The common lore is that inflation started when our universe exited the quantum gravity regime with 
energy densities \cite{Linde:2005ht,Linde:2007fr} 
\be
\label{equi}
\frac12 (\p \phi)^2 \lesssim V(\phi) \sim \frac12 \rho_{tot} \sim \frac12 M_P^4 . 
\ee
In this case the potential energy dominates the total energy density and the accelerated expansion starts 
even for a fundamentally small initial patch of Planck length radius $l_P$.  
The essential ingredient for  inflation in this setup is the existence of a nearly constant event horizon distance, 
also of size $\sim l_P$. The importance of the existence of the event horizon is that it protects the initial smooth patch 
from the outside inhomogeneities with nonzero gradients. 
If there was no event horizon, these inhomogeneities would infest the initial smooth patch 
and spoil inflation \cite{Goldwirth:1991rj}.

For the Starobinsky inflation we have 
\be
V_{inf} = \frac34 m^2 M_P^2 \sim 10^{-10} M_P^4 \ll M_P^4. 
\ee
For the total energy density when our universe exits the quantum gravity regime to be $\sim M_P^4$, 
one has to assume
\be
V(\phi) \ll \frac12 \dot \phi^2 \sim \rho_{tot}. 
\ee 
In other words, that a kinematic energy domination regime preceded the inflationary phase. 
In such a case,  $V(\phi)\ll \frac12 \dot{\phi}^2 \sim \rho_{tot}$, the scale factor grows like $t^{1/3}$ until the domination of the plateau potential yielding an event horizon of size 
\be
d_\text{event}(t \sim t_{P}) \sim 10^3 H^{-1}_P , 
\ee 
where $H^{-1}_P \equiv \sqrt{3}\, l_P$. 
Hence, one has to expel the density inhomogeneities at least $10^3$ Hubble scales further if the Universe has emerged from the Planck densities. 
The corresponding initially homogeneous volume is at least $10^9$ times bigger than $H_P^{-3}$ which means that, initially, one billion causally disconnected regions were much similar without any dynamical reason.

As we have seen the embedding of the Starobinsky model in the new-minimal supergravity framework 
has given rise to an additional propagating massive vector field. 
Pursuing a minimal setup it has been proposed in \cite{Dalianis:2015fpa} 
that the existence of this vector field can help to ameliorate the initial conditions problem. 
Indeed, 
one can accomplish an equipartition of the energy density 
\be
\frac12 \rho_{kin} \sim \frac12 \rho_{pot} \sim \frac12 M_P^4 , 
\ee 
by invoking non vanishing values not only for the scalar field but also for the components of the vector field ${\cal V}_m$.

\begin{figure}[t] 
\centering
\includegraphics [scale=.9, angle=0]{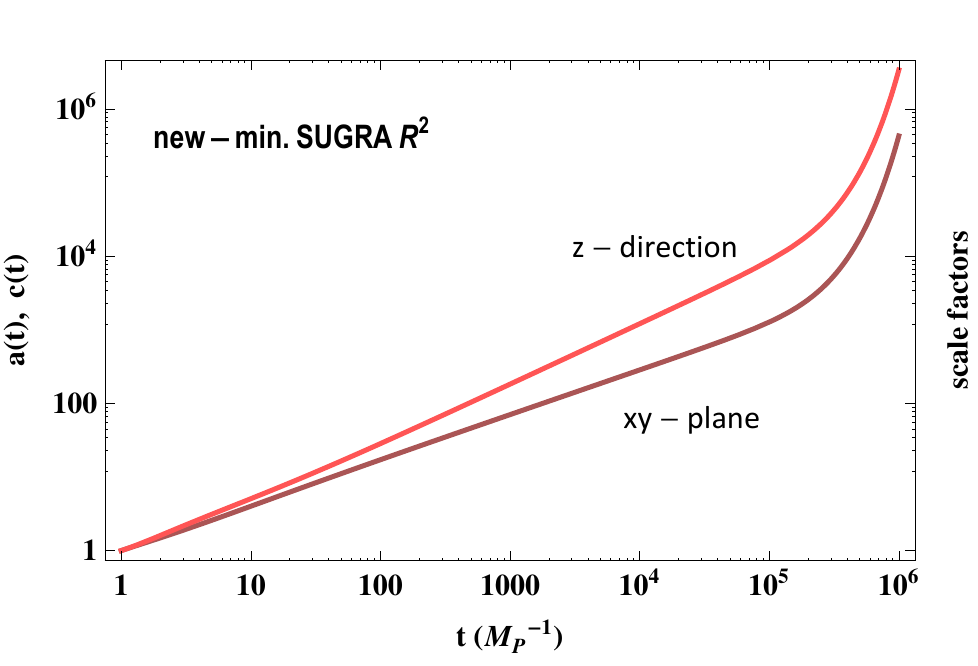} 
\caption{\small{The evolution of the scale factors. 
The anisotropic expansion of the Universe is manifest. 
After the onset of inflation $t_\text{INF}$ the scale factors evolve similarly and the anisotropy gets diluted. }}
\end{figure}

\begin{figure}[t] 
\centering
\includegraphics [scale=.9, angle=0]{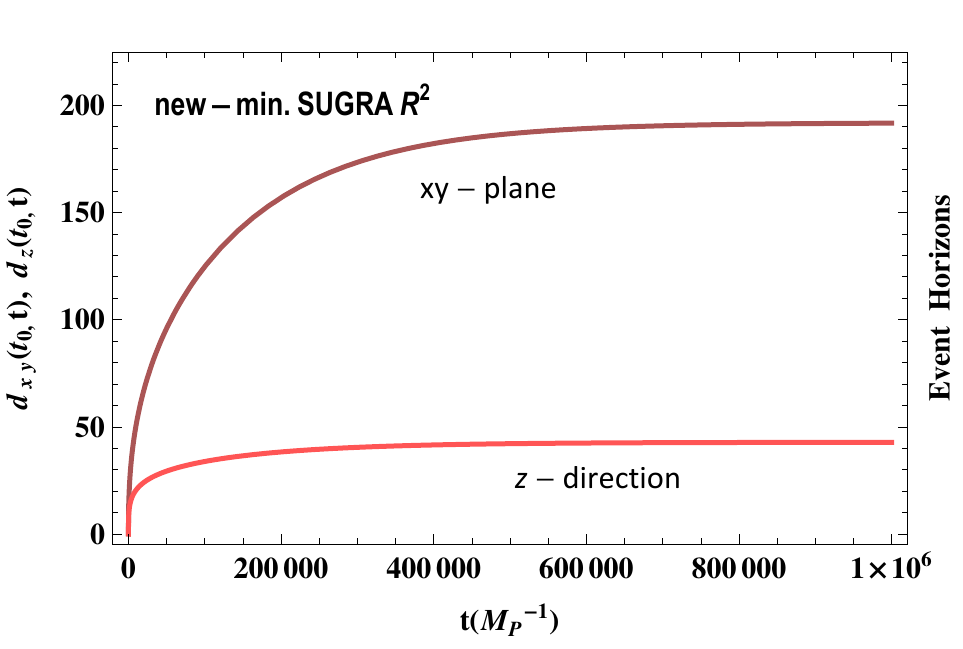}  
\caption{\small{The evolution of the  event horizons. 
The anisotropic expansion of the Universe  implies that the event horizon distance is respectively anisotropic. 
The event horizon distance is in $H^{-1}(t_P)=\sqrt{3} \, l_P$ units.}}
\end{figure}

One can choose the gauge 
\be
{\cal V}_0 &=&0 ,
\ee
and take the $z$-spatial axis  parallel to the direction of the vector 
\be
{\cal V}_i &=& {\cal A}_z(t) \delta_{i}^{z} . 
\ee
By giving to the vector a non-vanishing value a spatial direction is singled out which we have  identified  with the $z$-axis. 
This implies that the metric will be described by two scale factors 
\be \label{bianchi}
ds^2 = - dt^2 + a^2(t) [dx^2 + dy^2 ] +  c^2(t) dz^2 , 
\ee
hence, an anisotropy is created.

A numerical solution to the system of the equations and the evolution of the two scale factors, for $\rho_{kin, init}=0.5 V_{init}$ 
can be seen in  Figure 2 and Figure 3. Accordingly, the event horizon distances change in the $z$-direction and the $x-y$ plane. 
The initial condition problem is indeed  relaxed, but still a large homogeneous initial patch is required. 
For a complete analysis the reader is referred to \cite{Dalianis:2015fpa}, 
where also a discussion on the old-minimal supergravity embedding can be found.

It is worth mentioning that the initial conditions for Starobinsky inflation, 
in a pure gravitational  setup, need a minimum amount of  tuning for the case of open universe. 
Indeed there, the volume of the initial homogeneous patch is more than $10^6$ times smaller than 
the volume needed for closed or flat universe. For a complete discussion see \cite{Dalianis:2015fpa}.

\section*{Acknowledgments}

It is a great pleasure to thank I. Dalianis, A. Kehagias, A. Riotto and R. von Unge for our collaborations and discussions. 
This work was supported by the Grant agency of the Czech republic under the grant P201/12/G028.

\end{document}